\documentclass[aip,apl,superscriptaddress,reprint,floatfix,nobalancelastpage,amsmath,amsfonts,amssymb]{revtex4-1}
\usepackage{graphicx}

\begin{document}


\title{Photo-oxidative tuning of individual and coupled GaAs photonic crystal cavities} 



\author{Alexander Y. Piggott}
\email{piggott@stanford.edu}
\affiliation{E. L. Ginzton Laboratory, Stanford University, California, USA}
\author{Konstantinos G. Lagoudakis}
\affiliation{E. L. Ginzton Laboratory, Stanford University, California, USA}
\author{Tomas Sarmiento}
\affiliation{E. L. Ginzton Laboratory, Stanford University, California, USA}
\author{Michal Bajcsy}
\affiliation{E. L. Ginzton Laboratory, Stanford University, California, USA}
\affiliation{Institute for Quantum Computing, University of Waterloo, Ontario, Canada}
\author{Gary Shambat}
\affiliation{E. L. Ginzton Laboratory, Stanford University, California, USA}
\author{Jelena Vu\v{c}kovi\'{c}}
\affiliation{E. L. Ginzton Laboratory, Stanford University, California, USA}

\date{\today}

\begin{abstract}
We demonstrate a new photo-induced oxidation technique for tuning GaAs photonic crystal cavities using a $390~\mathrm{nm}$ pulsed laser with an average power of $10~\mathrm{\mu W}$. The laser oxidizes a small $\left(\sim 500~\mathrm{nm}\right)$ diameter spot, reducing the local index of refraction and blueshifting the cavity. The tuning progress can be actively monitored in real time. We also demonstrate tuning an individual cavity within a pair of proximity-coupled cavities, showing that this method can be used to correct undesired frequency shifts caused by fabrication imperfections in cavity arrays.
\end{abstract}

\pacs{}

\maketitle 

Photonic crystal cavities have received significant attention in recent years for their ability to strongly confine light with high quality factors \cite{yakahane_nature2003}. These unique attributes have enabled them to be used in extensive studies of cavity quantum-electrodynamics (QED) \cite{jvuckovic_prb2001,tyoshie_nature2004,denglund_nature2007}, and for the the implementation of a variety of compact classical devices such as low-power lasers \cite{bellis_np2014}, high-speed light-emitting diodes \cite{gshambat_nc2011}, and nonlinear frequency conversion devices \cite{krivoire_oe2009}. Many of these devices are fabricated using GaAs since thin membranes with embedded quantum dots or quantum wells can be grown epitaxially.

A number of exciting devices using coupled photonic crystal cavities have been proposed and demonstrated. In the cavity QED domain, a wide range of proposals using coupled photonic crystal cavities have been put forward, including sub-poissonian light generation \cite{tchliew_prl2010,mbamba_pra2011,amajumdar_prl2012}, the quantum simulation of exotic many-body systems \cite{adgreentree_np2006}, and quantum error correction \cite{jkerckoff_prl2010}. Coupled resonant oscillator waveguides (CROWs) can be constructed using a linear array of photonic crystal cavities \cite{ayariv_ol1999}. Coupled cavity array lasers have also been demonstrated \cite{haltug_oe2005}.

Coupled cavity devices require the cavity resonances to be spectrally aligned with each other. In addition, cavity QED devices generally require the cavity resonances to be aligned with the emitter. Due to fabrication imperfections, the resonant wavelength of identically designed photonic crystal cavities typically varies by several nanometers, even for devices only a few microns apart. Thus, there has been considerable interest in the post-fabrication tuning of photonic crystal cavities.

A number of techniques for post-fabrication tuning of GaAs photonic crystal cavities have been demonstrated. These include wet etching \cite{khennessy_apl2005}, infiltration of water \cite{nwlspeijcken_apl2012,svignolini_apl2010}, deposition of photosensitive materials \cite{tcai_apl2013,afaraon_apl2008} thermal oxidation \cite{hslee_apl2009}, atomic-force microscope (AFM) oxidation \cite{khennessy_apl2006}, green laser photo-oxidation \cite{fintonti_apl2012}, and application of strain to the entire chip \cite{ijluxmoore_apl2012}. However, many of these techniques are not well localized and hence cannot be used to tune individual cavities in coupled cavity configurations, while others require the application of fluids or polymers, or the use of an AFM.

We describe a new, more convenient technique for tuning GaAs photonic crystal cavities using $390~\mathrm{nm}$ pulsed laser light to introduce photo-induced oxidation. The laser oxidizes a small $\left(\sim 500~\mathrm{nm}\right)$ diameter spot, lowering the local index of refraction and blueshifting the cavity. Our approach exploits the same physical mechanism as Intonti et al.\cite{fintonti_apl2012}, which utilized a $532~\mathrm{nm}$ laser at a relatively high power ($700~\mathrm{\mu W}$). By using a shorter-wavelength laser, we were able to reduce the tuning power by nearly $2$ orders of magnitude while maintaining similar tuning rates, potentially enabling tuning of fragile structures such as nanobeam cavities \cite{ygong_oe2010}. Finally, to demonstrate the resolution and utility of our approach, we demonstrate tuning individual cavities in proximity-coupled pairs of cavities.

\begin{figure*}
 \includegraphics[height=9cm]{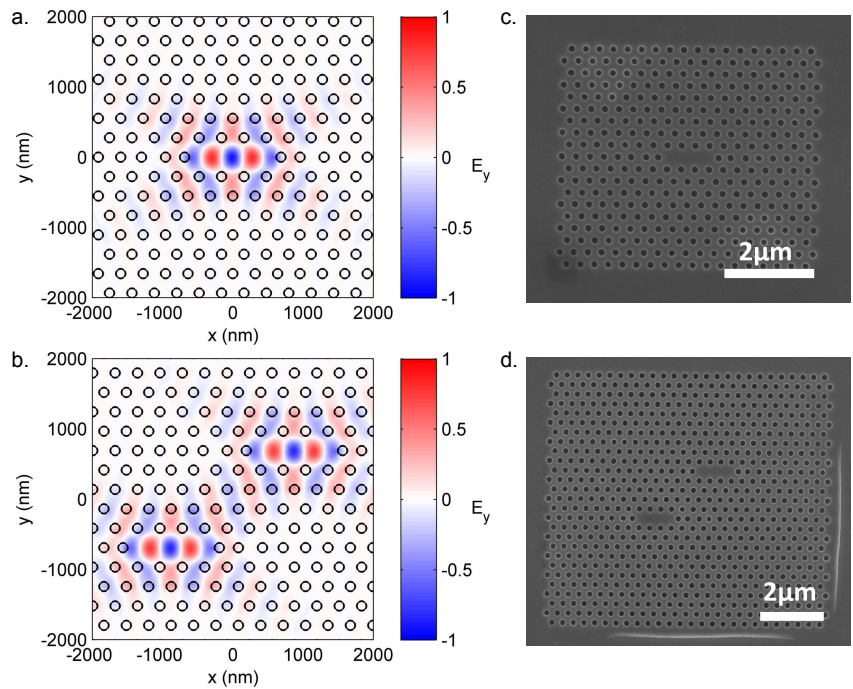}
 \caption{\label{fig:1} (a,b) The transverse electric field $(E_y)$ distribution for the fundamental modes of a (a) single L3 cavity and (b) two coupled L3 cavities, calculated in a finite-difference time domain (FDTD) simulation. The coupled cavity supports both anti-symmetric and symmetric modes; here we have plotted the latter. The black circles indicate the locations of the holes in the photonic crystal membrane. (c,d) Scanning electron microscopy (SEM) images of a single (c) and coupled (d) GaAs L3 cavities, taken before performing any tuning.}
\end{figure*}

The photonic crystal cavities used in this experiment were L3 cavities in a triangular photonic crystal lattice \cite{yakahane_nature2003}, with lattice constant $a = 336~\mathrm{nm}$ and design hole radius $r = 0.212a$. The fundamental mode for the L3 cavity calculated using finite-difference time-domain (FDTD) simulations is plotted in figure \textbf{\ref{fig:1}a}, and has a simulated quality factor of $\sim 4 \times 10^4$. We tested the proposed tuning mechanism on both individual cavities and pairs of proximity-coupled cavities. The coupled L3 defects were placed $5$ lattice periods apart, with a spectral splitting of $1.2~\mathrm{nm}$ calculated using FDTD. Scanning electron microscopy (SEM) images of these structures are shown in figure \textbf{\ref{fig:1}c} and figure \textbf{\ref{fig:1}d}.

The photonic crystal cavities were fabricated from GaAs wafers grown using molecular beam epitaxy, as described in previous works \cite{bellis_np2014}. The material stack consisted of a $220~\mathrm{nm}$ $\mathrm{GaAs}$ membrane and a $1500~\mathrm{nm}$ $\mathrm{Al_{0.8} Ga_{0.2} As}$ sacrificial layer on top of a GaAs substrate. The GaAs membrane contained $3$ layers of high-density InAs quantum dots $\left(300~\mathrm{dots / \mu m^2}\right)$ emitting at wavelengths near $1300~\mathrm{nm}$. The photonic crystal cavities were fabricated using electron-beam lithography, inductively-coupled plasma reactive-ion etching (ICP-RIE), and a final HF acid undercutting step, as described previously \cite{denglund_nature2007}.


The tuning was performed in a custom confocal microscopy setup coupled to a grating spectrometer with an InGaAs $\left(1.7~\mathrm{\mu m}\right)$ linear photodiode array. A Carl Zeiss LD-Plan-Neofluar $63\mathrm{x} / 0.75~\mathrm{Korr}$ was used as the microscope objective. A charge-coupled device (CCD) integrated into the microscopy setup was used to both image the sample, and determine where the lasers were focused. The experiments were all performed at room temperature in air, with the exception of an additional control test where the sample was placed in vacuum.

The photonic crystal cavities were tuned by simultaneously irradiating the sample with two lasers through the objective: the $390~\mathrm{nm}$ ultraviolet (UV) tuning laser, and an $830~\mathrm{nm}$ near-infrared pump laser to produce photoluminescence (PL) from the quantum dots embedded in the photonic crystal membranes.

For the tuning laser, we used a frequency-doubled pulsed Ti:Sapphire laser, producing an output wavelength of $390~\mathrm{nm}$, pulse repetition frequency of $80~\mathrm{MHz}$, and an average power of $10~\mathrm{\mu W}$ before the microscope objective. The pulse length was approximately $10~\mathrm{ps}$ after passing through a single-mode fiber (SMF) to clean up the beam profile. The UV laser was focused either directly on or immediately adjacent to the cavity to be tuned. The spot size of the tuning laser was roughly $500 - 700~\mathrm{nm}$ as estimated from SEMs of tuned devices.

An $830~\mathrm{nm}$, $350~\mathrm{\mu W}$ SMF-coupled continuous-wave multimode diode laser was used as the PL excitation laser. The PL laser was somewhat defocused in order to tightly focus the UV tuning laser, so it was necessary to use a relatively high power to produce a bright PL signal.  Due to the Purcell effect, the spontaneous emission rate from a photonic crystal cavity is strongly enhanced at its resonant frequencies \cite{empurcell_pr1946}. The photoluminescence spectrum was thus used to continuously monitor the cavity resonance during the tuning process. In principle, cross-polarized reflectivity measurements could also be used to monitor the cavity resonance when tuning non-photoluminescent devices \cite{denglund_nature2007}.

\begin{figure*}
 \includegraphics[height=9cm]{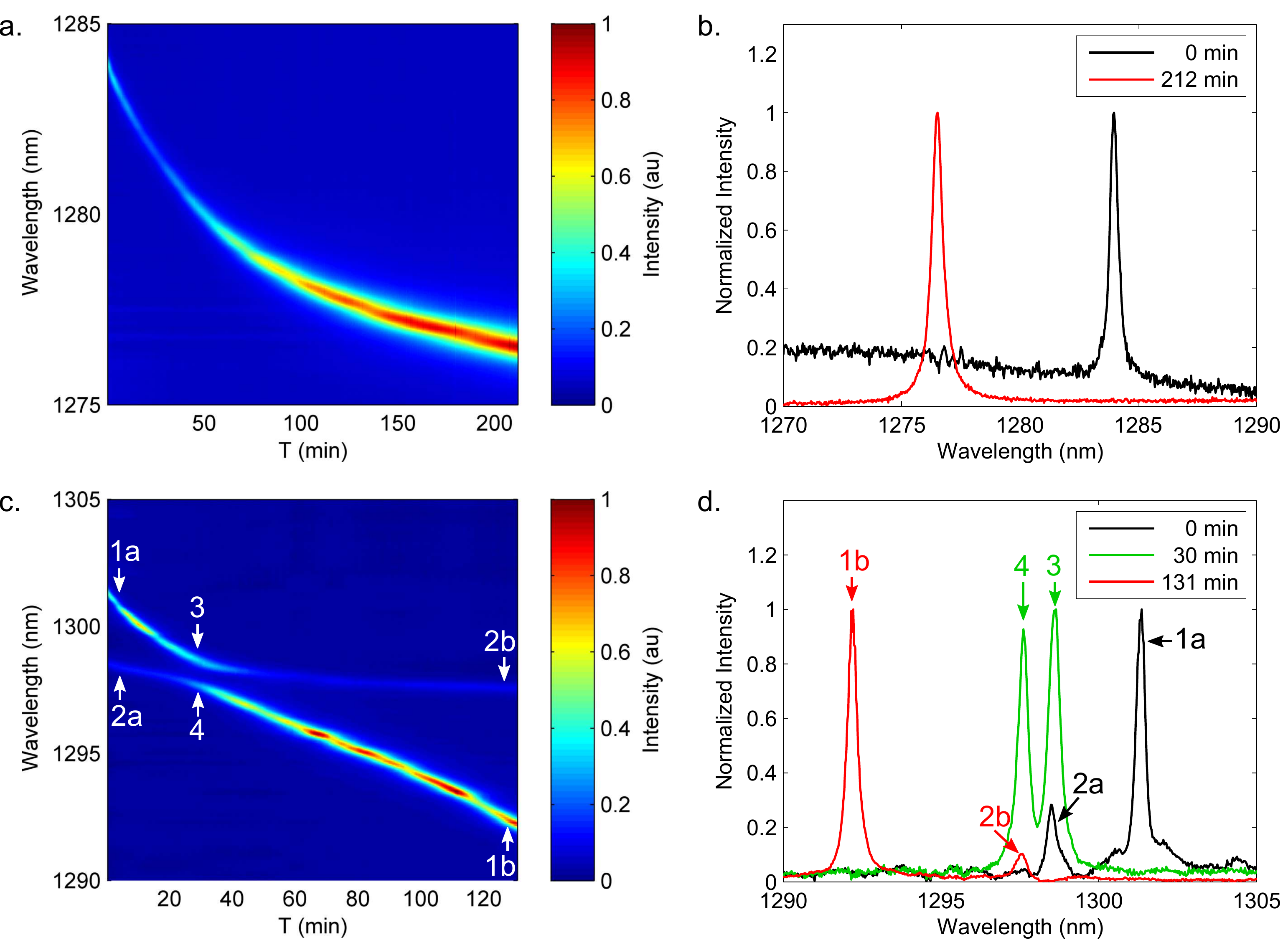}
 \caption{\label{fig:2} Photoluminescence spectrum of (a,b) a single L3 cavity and (c,d) coupled L3 cavities as a function of time while tuning with the $390~\mathrm{nm}$ UV laser. Due to Purcell enhancement, the cavity resonances are clearly visible. We have performed background subtraction for clarity.
(a) The single cavity was blueshifted by $7.8~\mathrm{nm}$ during the tuning process. (b) Initial and final photoluminescence spectra for the same cavity. The cavity quality factor was somewhat degraded by the tuning process, being reduced from $Q_{initial} = 4360$ to $Q_{final} = 2300$. 
(c) In the coupled-cavity system, one cavity was tuned by $9.1~\mathrm{nm}$, and the other cavity was tuned by only $1.0~\mathrm{nm}$, resulting in a clear anti-crossing where their resonances became degenerate. (d) Initial, intermediate (during the anticrossing), and final spectra for the coupled cavity system. The microscope was focused on the tuned cavity, resulting in a brighter PL signal from the tuned cavity than the untuned cavity.}
\end{figure*}

In figure \textbf{\ref{fig:2}a}, we present the tuning profile of a single L3 cavity. The tuning rate decreases as a function of time, suggesting a self-limiting mechanism. The tuning rate can be increased by increasing the UV laser power, but using excessively high power risks damaging the membrane due to thermal effects. The initial and final spectra are plotted in figure \textbf{\ref{fig:2}b}. The cavity quality factor $(Q)$ is somewhat degraded by the tuning process, being reduced from $Q_{initial} = 4360$ to $Q_{final} = 2300$.

Next, in figures \textbf{\ref{fig:2}c} and \textbf{\ref{fig:2}d} we present the tuning of a single cavity in a proximity-coupled pair of L3 cavities. The behaviour of such a system can be accurately described using coupled-mode theory \cite{hahaus_ieee1991}. Due to the coupling between the cavities, such a system will present two resonant peaks with frequencies $\Omega_1, \Omega_2$ given by
\begin{align}
\Omega_{1,2} = \frac{1}{2}\left(\omega_1 + \omega_2\right) \pm  \frac{1}{2}\sqrt{\left(\omega_1 - \omega_2\right)^2 + J^2}
\label{eqn:coupledcavityfreqs}
\end{align}
where $\omega_1,\omega_2$ are the individual cavity frequencies, and $J$ is the coupling between the cavities. We have assumed the cavities are in the strong coupling limit $J \gg \omega_i / Q_i$, where the $Q_i$ is the quality factor of cavity $i$.

\begin{figure}
 \includegraphics[width=6.5cm]{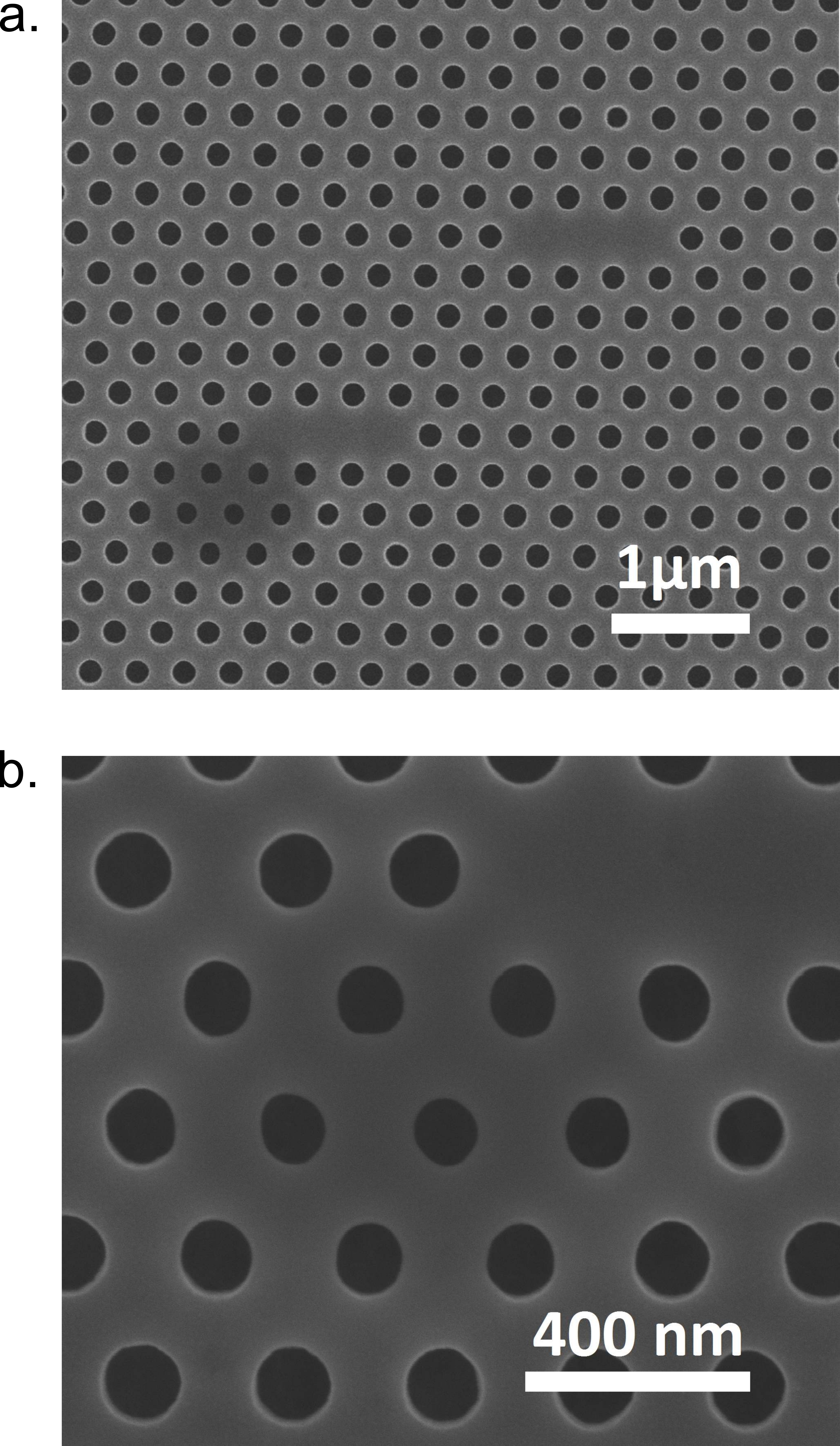}
  \caption{Scanning electron microscopy (SEM) image of a coupled L3 system after performing tuning, showing (a) the entire structure and (b) zoomed in on the laser oxidized spot. The oxidation is visible as a slight discoloration, and the photonic crystal holes are also reduced in size due to the growth of oxide on the surface. \label{fig:3}}
\end{figure}

The UV laser was focused on one edge of a cavity, as can be seen in an SEM of the tuned structure in figure \textbf{\ref{fig:3}a}. As the UV laser was applied, the resonant peak at $1298.5~\mathrm{nm}$ remained nearly stationary while the other resonant peak blueshifted from $1301.3~\mathrm{nm}$ to $1292.2~\mathrm{nm}$. As the two peaks pass each other, a clear anti-crossing - which arises from equation (\ref{eqn:coupledcavityfreqs}) - can be observed.

The tuning mechanism is likely photo-induced oxidation of GaAs by the $390~\mathrm{nm}$ UV laser, resulting in reduction of the local index of refraction and blueshifting the cavities. Previous research has shown that photo-oxidation of GaAs surfaces can be induced by UV irradiation under similar parameters to our experiment \cite{cfyu_cpl1986,cfyu_jvstb1987,zlu_jcp1990}. SEMs of tuned cavities are shown in figure \textbf{\ref{fig:3}}. No damage to the photonic crystal is visible other than a slight discoloration and reduction in hole size in the vicinity of the irradiated spot, probably due to the growth of oxide on the surface. We also conducted identical experiments in a vacuum chamber pumped down to $\sim 10^{-4}~\mathrm{Torr}$, and there was no observable tuning or change in appearance.

Due to the low power of our UV tuning laser, the tuning mechanism is very unlikely to be thermal oxidation. The steady state temperature increase should be very small. Based on Sentaurus simulations of similar structures, we expect to see a temperature rise of $< 1~\mathrm{K}$ for a heat dissipation of $10~\mathrm{\mu W}$ \cite{jpetykiewicz_apl2012}. We also see no permanent tuning effects from our higher power $350~\mathrm{\mu W}$ PL laser.

The instantaneous temperature rise from each UV laser pulse is expected to be much higher, but still relatively low. Assuming a plane wave propagating into an infinite slab of GaAs, and ignoring reflections, the local temperature rise $\Delta T$ from a single pulse is given by
\begin{align}
\Delta T = \Phi_0 \frac{\alpha e^{- \alpha z}}{\rho C}
\label{eqn:laserpulseheating}
\end{align}
where $\Phi_0$ is the incident fluence $\left(\mathrm{W / cm^2}\right)$, $\alpha = 7.433\times10^5~1/\mathrm{cm}$ is the extinction coefficient of GaAs at $390~\mathrm{nm}$ \cite{edpalik_hoc1997}, $C = 0.350~\mathrm{J / g\; K}$  and $\rho = 5.320~\mathrm{g / cm^3}$ are the heat capacity and density of GaAs \cite{smsze_scsensors1994}, and $z$ is the distance from the incident surface. If we assume the incident light is a gaussian beam with a diameter of $500~\mathrm{nm}$, the estimated temperature rise at the surface $\left( z = 0 \right)$ is $50.8~\mathrm{K}$, far too low for thermally-induced oxidation. Since we assumed there are no reflections, and used a conservative estimate of laser spot size based on our SEMs, this should be an overestimate of the actual temperature rise.

In conclusion, we have demonstrated a technique for tuning GaAs nanophotonic resonators which requires only a low-power UV laser at room temperature in ambient atmosphere. In particular, this technique can be used to independently tune individual cavities in proximity-coupled cavity configurations, allowing the fabrication of a wide range of coupled-cavity devices with applications ranging from quantum simulations and information processing to low-power nanophotonic lasers.

\begin{acknowledgments}
The authors acknowledge support provided by the Air Force Office of Scientific Research (AFOSR) MURI Center for Multi-Functional Light-Matter Interfaces based on Atoms and Solids (FA9550-09-1-0704), and the AFOSR MURI for Complex and Robust On-chip Nanophotonics (FA9550-12-1-0025). AYP acknowledges support from the Stanford Graduate Fellowship. KGL acknowledges support from the Swiss National Science Foundation.
\end{acknowledgments}



%
%

%



\bibliography{UV_photoox_references}

\end{document}